\documentclass[runningheads]{llncs}

\usepackage{graphicx}
\newcommand{\orcid}[1]{\href{https://orcid.org/#1}{\textcolor[HTML]{A6CE39}{\aiOrcid}}}

\usepackage{algorithm}
\usepackage{algpseudocode}
\usepackage{orcidlink}
\usepackage{makecell}
\usepackage{array}

\title{Securing Tomorrow's Smart Cities: Investigating Software Security in Internet of Vehicles and Deep Learning Technologies}
\author{ Ridhi Jain\orcidlink{0000-0002-6102-7114} \and Norbert Tihanyi\orcidlink{0000-0002-9002-5935} \and Mohamed Amine Ferrag\orcidlink{0000-0002-0632-3172}}
\institute{
Technology Innovation Institute\\
 		\email{\{ridhi.jain,norbert.tihany,mohamed.ferrag\}@tii.ae}}
\date{}

\begin{document}
\maketitle              % typeset the header of the contribution
\begin{abstract}

Integrating Deep Learning (DL) techniques in the Internet of Vehicles (IoV) introduces many security challenges and issues that require thorough examination. This literature review delves into the inherent vulnerabilities and risks associated with DL in IoV systems, shedding light on the multifaceted nature of security threats. Through an extensive analysis of existing research, we explore potential threats posed by DL algorithms, including adversarial attacks, data privacy breaches, and model poisoning. Additionally, we investigate the impact of DL on critical aspects of IoV security, such as intrusion detection, anomaly detection, and secure communication protocols. Our review emphasizes the complexities of ensuring the robustness, reliability, and trustworthiness of DL-based IoV systems, given the dynamic and interconnected nature of vehicular networks. Furthermore, we discuss the need for novel security solutions tailored to address these challenges effectively and enhance the security posture of DL-enabled IoV environments. 
% By synthesizing insights from existing literature, this review aims to provide a comprehensive understanding of the security landscape surrounding DL techniques in the context of the IoV, paving the way for future research and innovation in this critical domain. 
By offering insights into these critical issues, this chapter aims to stimulate further research, innovation, and collaboration in securing DL techniques within the context of the IoV, thereby fostering a safer and more resilient future for vehicular communication and connectivity.

\end{abstract}
%
%
%

%%%%%%%%%%%%%%%%%%%%%%%%%%%%%%%%%%%%%%%%%%%%%%%%%%%%%%%%%%%%%%%%%%%%
\section{Introduction}

\begin{figure*}[t]{\vspace{0pt}}{\vspace{0pt}}
\centering
\includegraphics[width=0.8\textwidth]{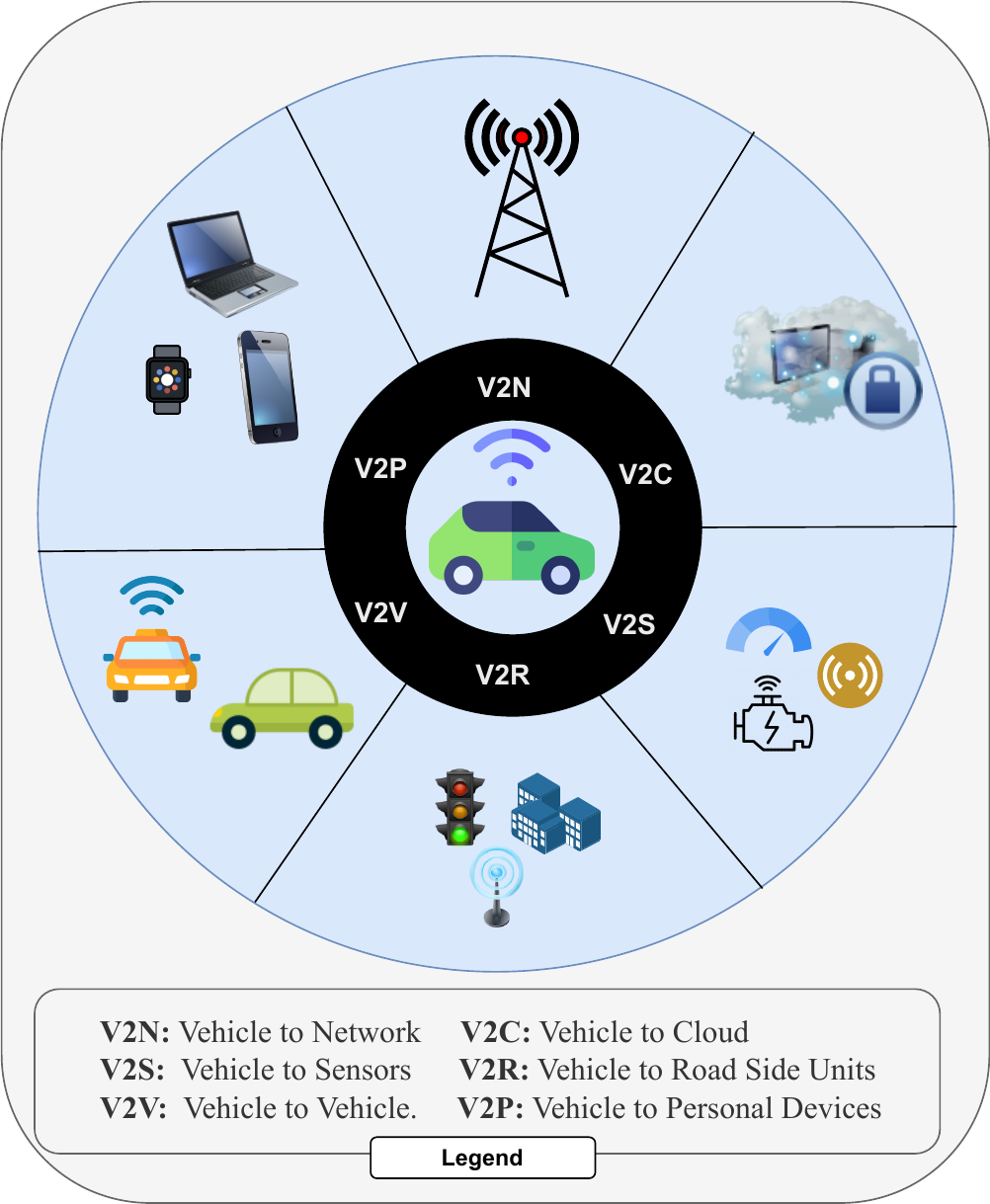}
\caption{Different types of communications in Internet of Vehicles (IoV)}
\label{fig:communication}
\end{figure*}

Recent projections indicate a significant surge in the automotive industry, with estimates suggesting that the global vehicle count will reach two billion by 2030~\cite{sperling2009two}, \cite{gross2016planet}. As the automotive landscape continues to evolve, the autonomous vehicles market is predicted to skyrocket from $230,000$ in 2025 to $11.8$ million in 2035. Most vehicles in use are expected to be autonomous by 2050~\cite{hendrickson2014connected}. This exponential growth in autonomous vehicle adoption is mirrored by the projections for the global market size of smart connected cars, which is anticipated to reach trillions of dollars by 2035~\cite{kong2021study}. 
% Recent studies indicate that this planet will have two billion cars by 2030~\cite{sperling2009two},\cite{gross2016planet}.
However, as the number of connected vehicles increases, ensuring robust software security becomes paramount to safeguarding these systems' integrity, privacy, and safety. 

\begin{figure*}[t]{\vspace{0pt}}{\vspace{0pt}}
\centering
\includegraphics[width=1\textwidth]{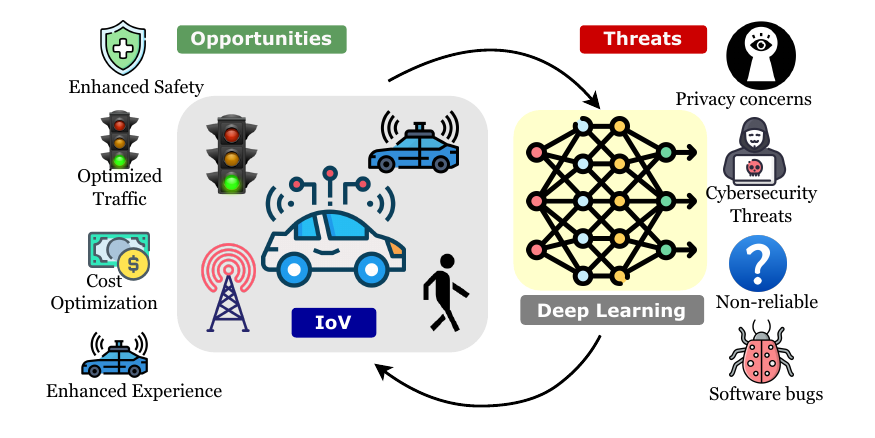}
\caption{Threats and opportunities in integrating Internet of Vehicles (IoV) with Deep Learning (DL)}
\label{fig:threats-opp}
\end{figure*}
As urban centers embrace smart cities, the convergence of IoT technologies and transportation systems presents new opportunities and challenges (see Figure~\ref{fig:threats-opp}). Integrating the Internet of Vehicles (IoV) allows vehicles to communicate with each other and the surrounding infrastructure, optimizing traffic flow, enhancing safety, and reducing environmental impact, as shown in Figure~\ref{fig:communication}. However, this interconnectedness also increases cyber threat targets. Ensuring privacy and compliance with data protection standards is crucial as smart cities collect vast amounts of personal data. The extensive IoV network is vulnerable to cyberattacks like intrusion, data breaches, and malicious manipulation of vehicle systems. Wireless communication introduces additional risks, such as eavesdropping and spoofing. Thus, robust security measures are essential to protect the integrity, confidentiality, and availability of communication and data exchange within the IoV ecosystem.

The advent of Deep Learning (DL) technologies enhances IoV capabilities, from predictive maintenance to autonomous driving. By leveraging deep neural networks, IoV systems can improve vehicle safety with real-time collision detection and avoidance mechanisms~\cite{chang2019deepcrash,almutairi2023hybrid,wang2020collision}, optimize traffic flow by analyzing patterns and suggesting efficient routes~\cite{hussain2021artificial}, and enable autonomous driving through advanced environment perception and decision-making~\cite{lu2019cognitive}. Additionally, DL facilitates predictive maintenance~\cite{theissler2021predictive} by analyzing sensor data to anticipate failures and perform proactive servicing while enabling driver behavior analysis to assess attentiveness and predict safety hazards.

Besides enhancing the IoV capabilities, DL techniques can contribute to software security in IoV by facilitating intrusion detection~\cite{kang2016intrusion}, \cite{ullah2022hdl}, \cite{ahmed2021deep} and anomaly detection~\cite{garg2019sec}, \cite{aziz2022anomaly} in vehicular networks. By analyzing network traffic patterns and identifying unusual behaviors, DL models can detect cyber threats~\cite{aslam2024securing}, such as malicious intrusions or unauthorized access attempts, and trigger appropriate security responses to mitigate potential risks. Moreover, DL-based techniques can verify the integrity and security of firmware and software components in connected vehicles, ensuring they are free from vulnerabilities and malware.

Although DL promises to revolutionize the IoV, the widespread adoption of these technologies also introduces new challenges, particularly concerning software security.
% Security issues and challenges inherent to deep learning (DL) techniques in the Internet of Vehicles (IoV) stem from several factors. Firstly, 
DL models are vulnerable to adversarial attacks~\cite{ji2018model}, \cite{zhang2020adversarial}, \cite{shafee2021privacy}, where malicious actors can manipulate input data to deceive the model and produce incorrect outputs, leading to safety risks in vehicular systems. Additionally, DL models often rely on large datasets for training, raising concerns about data privacy and confidentiality~\cite{bae2018security}, \cite{hitaj2017deep}, especially when sensitive vehicular data is involved. DL models may suffer from model poisoning attacks, where adversaries inject malicious data into the training dataset to compromise the model's performance or integrity~\cite{yu2020cloudleak}, \cite{kreuk2018deceiving}. Moreover, the complexity and opacity of DL models make them difficult to interpret and understand, hindering efforts to verify their security and trustworthiness in IoV applications. Further, DL models are susceptible to evasion attacks~\cite{jiang2020poisoning}, \cite{ayub2020model}, where adversaries exploit vulnerabilities in the model's decision boundary to evade detection or classification, posing risks to the security of IoV systems. 
Addressing these security challenges requires robust security measures, including adversarial robustness techniques, data privacy mechanisms, model verification methods, and explainable AI approaches tailored to the unique requirements of IoV environments.
% Nevertheless, these DL techniques come with their own challenges.

% The sale of autonomous vehicles is predicted to increase from $230,000$ in 2025 to 11.8 million in 2035. Most vehicles in use will be autonomous by 2050~\cite{hendrickson2014connected}.  

% The global market size of smart connected cars is predicted to reach trillions of dollars in 2035~\cite{kong2021study}.

% The concept of smart cities, characterized by integrating advanced technologies to enhance urban living, has gained significant traction in recent years. Central to this vision is Internet of Things (IoT) devices, particularly Internet of Vehicles (IoV) and computer vision technologies, which promise to revolutionize transportation, infrastructure management, and public safety. However, the widespread adoption of these technologies also introduces new challenges, particularly concerning software security.

As we explore the intersection of DL technologies and IoT, it's crucial to understand the challenges, opportunities, and implications.
This chapter comprehensively explores software security in the context of the IoV, intersecting with the transformative potential of DL technologies. We examine common vulnerabilities, attack vectors, and security mechanisms relevant to these technologies, drawing insights from existing literature. Through a multifaceted investigation, we aim to shed light on the critical aspects of software security within IoV ecosystems and pave the way for informed strategies to address cybersecurity threats and foster the sustainable development of secure urban environments.

%%%%%%%%%%%%%%%%%%%%%%%%%%%%%%%%%%%%%%%%%%%%%%%%%%%%%%%%%%

\section{Related Work}

% \subsection{Internet of Vehicles (IoV) Advancements}
 % Another critical area of development has been in the integration of battery technology, propulsion systems, power interfaces, and vehicle network systems for autonomous and connected electric vehicles~\cite{abro2023comprehensive},\cite{islam2023advancements}. 
 The IoV has gained enormous popularity with the advancements in vehicle connectivity and communication technologies, as they have enabled the integration of vehicles with the internet, allowing for real-time data exchange and intelligent decision-making.
 Recent works have focused on optimizing the IoV, such as the integration of battery technology, propulsion systems, power interfaces, and vehicle network systems for autonomous and connected electric vehicles~\cite{abro2023comprehensive},\cite{islam2023advancements}. Another work by Mollah et al.~\cite{mollah2020blockchain} leverages blockchain for intelligent transport.

 Integrating DL technologies with IoV has sparked significant advancements, particularly in enhancing vehicular systems' capabilities for autonomous driving and traffic management. Prior works leverage real-time data from diverse sensors and vehicles, predict traffic congestion~\cite{akhtar2021review}, \cite{saleem2022smart}, \cite{leen2023mitigating}, and optimize traffic signals~\cite{ning2020joint}, \cite{chen2020traffic} to streamline flow and minimize waiting times. Similar techniques have also been proposed for accident detection~\cite{chang2019deepcrash}, \cite{zhao2022deep} and prevention~\cite{wang2020collision}.

 DL has also shown to improve IoV's capabilities in object detection and traffic pattern analysis~\cite{chen2020traffic}. Advanced neural networks, such as convolutional neural networks (CNNs)~\cite{lecun1995convolutional} and recurrent neural networks (RNNs)~\cite{medsker1999recurrent}, have been effectively used to process and interpret the continuous data stream from vehicle sensors and cameras. Mezair et al.~\cite{mezair2022advanced} an Advanced DL framework for the Internet of Behaviors (IoB) applied to connected vehicles, which integrates different DL models to handle heterogeneous sensor data and optimizes hyperparameters using a Branch-and-Bound strategy, resulting in superior performance in terms of accuracy and runtime for predicting traffic flow. Multi-Agent Reinforcement Learning (MARL) has been used to enhance the control and coordination of connected and automated vehicles~\cite{hua2023multi},\cite{yadav2023comprehensive}. Chen et al.~\cite{chen2016vision} introduce a vision-based ego-positioning system within the IoV architecture. This system employs a unique weighted k-cover technique to compress 3D point cloud models, thereby retaining crucial structural details. Computer vision-powered IoT devices, like smart surveillance cameras~\cite{rohith2021surveillance} and agricultural drones~\cite{wei2019spatial}, have wide-ranging applications. 
 Additionally, progress in federated learning and edge computing enables these devices to learn together from their surroundings while keeping sensitive visual data private~\cite{kong2021federated}, \cite{singh2021federated}, \cite{ning2019mobile}, \cite{ning2019mobile}.

 As the high interconnectedness of IoV makes them immensely susceptible to security attacks, many recent works focus on mitigating them. Researchers explore 6G enabled IoV for enhanced security and privacy~\cite{osorio2022towards}, \cite{xu2021service}, \cite{xu2022secure}. Nie et al.~\cite{nie2020data} introduce a data-centric approach to Intrusion Detection Systems (IDS) by examining the link load behaviors of Road Side Units (RSUs) in the IoV to identify anomalies caused by different attacks. Given the severity of these attacks, many similar works propose IDS~\cite{kang2016intrusion}, \cite{aljarrah2019intrusion}. 
Further efforts have been geared towards creating robust defensive mechanisms against various cyber attacks~\cite{yue2024internet}, including spoofing~\cite{oligeri2022gps} and denial of service (DoS) attacks~\cite{tian2019vcash}, \cite{sedar2022multi}, to address the unique challenges posed by the IoV's dynamic and distributed nature.

The increasing reliance on DL and AI-driven applications in the IoV has exposed the IoV to their vulnerabilities. One significant concern is the susceptibility of DL models to adversarial attacks~\cite{zhang2020adversarial}, \cite{yu2020cloudleak}, \cite{ayub2020model}. %, where malicious inputs designed to deceive AI models can lead to misinterpretations and erroneous actions by autonomous vehicles. 
% Research has highlighted the development of robust deep learning architectures and adversarial training techniques as critical measures to mitigate such risks.
Studies suggest the effectiveness of incorporating noise and perturbations during the training phase to enhance model resilience against such attacks~\cite{qiu2021security}, \cite{yang2021generating}.

 % One key development area has been applying deep learning for real-time vehicle diagnostics and predictive maintenance. These methods utilize vast datasets generated by IoV systems to predict vehicle malfunctions before they occur, significantly reducing downtime and maintenance costs.

Another important aspect is data security, which DL systems in IoV use. The integrity~\cite{shen2022innovative}, \cite{javed2020odpv}, and confidentiality~\cite{zavvos2021privacy} of the data collected from sensors and used for training and operational purposes are paramount. Encryption techniques~\cite{rathore2022novel}, \cite{wu2023cloud}, and secure data transmission protocols~\cite{chen2019secure} are being researched to protect data from interception and tampering. Furthermore, implementing federated learning, where data can be processed locally rather than sent to a centralized server, reduces the risk of data breaches while still allowing collective insights from multiple vehicles~\cite{lu2020blockchain}, \cite{chai2020hierarchical}.

\section{Deep learning in Internet of Vehicles}

\begin{table}[t]
    \centering
    \scriptsize
    % \addtolength{\tabcolsep}{-1pt}
% \setlength\extrarowheight{15pt}
% \setlength{\abovecaptionskip}{5pt}
% \setlength{\belowcaptionskip}{-10pt}
    \begin{tabular}{p{2.8cm}|p{1.7cm}|p{3cm}|p{4.2cm}}
    \hline
        \textbf{Application} & \makecell{\textbf{DL} \\ \textbf{Techniques}} & \textbf{Benefits} & \textbf{Example} \\
        \hline
        \hline
         Autonomous Driving \cite{chang2019deepcrash},~\cite{almutairi2023hybrid},~\cite{wang2020collision},~\cite{hussain2021artificial},~\cite{lu2019cognitive}& CNNs, RNNs, GANs & Enhanced perception,  decision making & Self-driving cars detecting pedestrians and obstacles in real-time \\
        \hline 
        
        Traffic Prediction \cite{hussain2021artificial}, ~\cite{akhtar2021review}, ~\cite{saleem2022smart}, ~\cite{leen2023mitigating}, ~\cite{ning2020joint}, ~\cite{chen2020traffic},~\cite{chang2019deepcrash}, ~\cite{zhao2022deep},~\cite{wang2020collision} & LSTM, GRU, RNNs & Improved traffic management and routing & Predicting traffic congestion and suggesting optimal routes\\
        \hline
        Vehicle-to-Vehicle (V2V) Communication ~\cite{dorner2017deep}, ~\cite{wang2021transfer}, ~\cite{suresh2021modeling} & DRL & Efficient and safe communication between vehicles & Vehicles coordinating to avoid collisions and optimize speed \\
        \hline
        Driver Behavior Analysis ~\cite{chhabra2023privacy}, ~\cite{chen2020driver}, ~\cite{li2022macroscopic} & CNNs,
        RNNs & Increased safety and personalized driving experience & Monitoring driver attention and detecting fatigue to prevent accidents \\
        \hline
        Predictive Maintenance ~\cite{theissler2021predictive}, ~\cite{yang2024prem} & DAE, CNNs & Reduced downtime and maintenance costs & Predicting vehicle component failures before they happen based on sensor data \\
        \hline
       % Smart Parking & \makecell[l]{CNNs\\ YOLO} & \makecell[l]{Efficient parking \\ space utilization} & \makecell[l]{Identifying available parking\\ spots and guiding vehicles \\to them using real-time \\ video analysis} \\
        % \hline
         Security in IoV  \cite{yang2022transfer}, ~\cite{nie2020data}, ~\cite{peng2019internet}, ~\cite{alferaidi2022distributed}, \cite{ahmed2021deep} & CNNs, LSTM & Improved resilience and robustness of vehicular networks & Intrusion section, malware detection and access control \\
        \hline
    \end{tabular}
    \caption{Usage of Deep Learning in Internet of Vehicles.
    \textbf{DAE:} Deep AutoEncoder, \textbf{DRL:} Deep Reinforcement Learning, \textbf{CNN:} Convolutional Neural Network, \textbf{RNN:} Recurrent Neural Network, \textbf{GAN:} Generative Adversarial Network, \textbf{YOLO:} You Only Look Once, \textbf{LSTM:}  Long Short-Term Memory
    }
    \label{table:deep_learning_iov}
\end{table}

Deep learning (DL) greatly enhances the IoV by enabling vehicles and infrastructure to process complex data streams efficiently (see Table~\ref{table:deep_learning_iov}). IoV uses DL for tasks like image recognition~\cite{islam2018image} and decision-making~\cite{asmika2021decision}, and improving safety~\cite{peng2018safety} by detecting and mitigating potential hazards on the road. DL models can be trained to predict traffic congestion patterns~\cite{saleem2022smart}, \cite{akhtar2021review}, \cite{leen2023mitigating} and optimize route planning~\cite{lin2018route}. 
Deep learning supports autonomous driving systems by helping vehicles understand their surroundings and make driving decisions. Vehicles learn optimal driving strategies through deep reinforcement learning, continually improving performance in various road conditions. 

\subsection{Convolutional Neural Networks (CNNs) in Internet of Vehicles}

% Neural networks are a fundamental component of the burgeoning field of the Internet of Vehicles (IoV), offering advanced solutions for processing vast amounts of data generated by connected vehicles. 
As a subset of machine learning, neural networks learn to perform tasks by considering examples, generally without task-specific programming. In the context of IoV, they enable a wide range of applications, from autonomous driving to traffic management, enhancing the efficiency and safety of modern transport systems. These networks process inputs from various sensors and cameras installed in the vehicle, such as LIDAR, radar, and optical cameras, and make sense of complex visual environments to navigate safely. They help detect objects, interpret road signs, and make split-second decisions about vehicle maneuvering, mimicking human cognitive driving abilities.

Neural networks, particularly CNNs for visual data and RNNs excel in the IoV due to their proficiency in pattern recognition and temporal modeling. CNNs, a popular DL tool, analyze visual data from vehicle cameras, aiding in lane detection and pedestrian recognition~\cite{li2017scale}. Besides advancing the functionalities of the IoV, CNNs can also enhance security in such frameworks, like providing IDS~\cite{yang2022transfer},~\cite{nie2020data},~\cite{peng2019internet},~\cite{alferaidi2022distributed}, \cite{ahmed2021deep} traffic accident prediction~\cite{zhao2022deep}~\cite{chang2019deepcrash},~\cite{lin2021intelligent}.

In traffic management, neural networks optimize flow and reduce congestion by analyzing real-time data from vehicle-to-vehicle and vehicle-to-infrastructure communications. They learn from historical traffic patterns to predict volumes and suggest the best routes, adapting to conditions like accidents or roadworks. Neural networks also enhance vehicle safety through predictive analytics and real-time monitoring, predicting potential component failures from diagnostic data to enable preemptive maintenance and prevent accidents. Additionally, they detect and alert drivers to road hazards, such as unexpected pedestrian movements or sudden obstacles.

% \end{enumerate}

\subsection{Recurrent Neural Networks (RNNs) in Internet of Vehicles}

Recurrent neural networks (RNNs) and its variant, Long Short-Term Memory (LSTM) ~\cite{hochreiter1997long} networks handle sequential data tasks, like predicting traffic flow patterns~\cite{fu2016using}. Their adaptability allows them to adjust to changing road conditions and traffic patterns, while their real-time decision-making capabilities ensure timely responses. Moreover, neural networks scale efficiently to handle large-scale IoV deployments, processing vast amounts of data from millions of connected vehicles and infrastructure components to provide actionable insights and support intelligent decision-making, making them integral to the success of IoV ecosystems. IoV systems optimize maintenance operations~\cite{singh2021federated},~\cite{atallah2018scheduling}, reduce downtime, and extend vehicle lifespan~\cite{huo2022lifespan},~\cite{rao2023optimizing} using DL. This proactive approach enhances safety and reliability \cite{sharma2018security}, benefiting both individual vehicle owners and fleet operators.

LSTM networks, a type of  RNN, are particularly valuable in the IoV for handling sequential data. They excel in processing time-series data from vehicle sensors, enabling predictive analytics and real-time decision-making essential for dynamic vehicle operations. LSTM networks effectively predict traffic conditions by learning from historical data, aiding in route planning and navigation to reduce congestion and improve fuel efficiency. Additionally, LSTM networks analyze vehicle telematics data to predict breakdowns, allowing timely maintenance actions to enhance vehicle longevity. In autonomous driving, LSTMs process sequential sensor data for dynamic decision-making, ensuring safety by understanding other vehicles' behavior and predicting potential hazards like lane changes or sudden stops. LSTMs can predict safety threats by analyzing historical accident data and triggering automatic safety measures to prevent incidents. They also process real-time data from sensors and V2X communication for instant decision-making, which is crucial for adapting to changing conditions or responding to emergencies.

\subsection{Generative Adversarial Networks (GANs) in Internet of Vehicles}

Generative Adversarial Networks (GANs) offer significant potential for enhancing various aspects of the IoV. A GAN is a DL model with two neural networks, a generator and a discriminator, competing to create realistic data. In IoV applications, GANs play a crucial role in generating realistic synthetic images of traffic scenes, which improves training data quality for tasks such as object detection. Additionally, GANs contribute to IoV security by assisting in anomaly detection to identify irregularities in vehicle behavior. Moreover, they aid in diversifying datasets for training DL models, particularly when acquiring real-world data is challenging, enhancing the robustness and efficiency of IoV applications. 

GANs are especially valuable for data augmentation in autonomous driving scenarios, where they generate synthetic data resembling real-world conditions, accelerating model training and ensuring readiness for diverse road situations. Furthermore, GANs facilitate the creation of realistic simulations for testing and development purposes in IoV systems, allowing comprehensive evaluation of vehicle responses in dynamic yet controlled environments. Security-wise, GANs play a critical role in strengthening IoV security frameworks by generating synthetic attack data for training detection systems, enabling them to recognize better and respond to real threats. Additionally, GANs contribute to anomaly detection by learning normal behavior patterns, enabling the identification of deviations indicative of potential security breaches or system faults.

\subsection{Large Language Models (LLMs) in Internet of Vehicles}

Large Language Models (LLMs) are increasingly being integrated into the IoV ecosystem to enhance communication, interaction~\cite{cui2024drive}, and decision-making processes~\cite{cui2024receive}. 
In the IoV, LLMs leverage natural language processing (NLP) techniques to interpret and generate human-like text, enabling various applications such as voice-enabled assistants, human-vehicle interactions, and contextual understanding of traffic-related information. These models allow vehicles to comprehend verbal commands from drivers, provide real-time updates on road conditions, and communicate with other vehicles and infrastructure components. Additionally, LLMs empower IoV systems with advanced capabilities like semantic analysis of traffic reports, sentiment analysis of social media data, and personalized recommendations for navigation and entertainment, enhancing user experiences, safety, and operational efficiency. 

Integrating LLMs into vehicle systems makes interactions more intuitive and conversational, enabling dynamic feedback and providing actionable insights from textual data generated by IoV systems. LLMs also contribute to safety features by processing emergency data and facilitating communication in emergencies, extending beyond vehicle-to-driver interactions to vehicle-to-vehicle (V2V) and vehicle-to-infrastructure (V2I) communication, where they assist in managing and translating data into actionable vehicle advice.

\section{Security Threats and Limitations}

\subsection{Software Security in Deep Learning Technologies}

Deep Learning technologies are crucial in modern smart cities, providing surveillance, traffic management, object recognition, and augmented reality solutions. Despite their benefits in efficiency and automation, they pose security challenges that require attention to protect privacy~\cite{wang2019privacy}, integrity, and trust in the systems.

DL systems are vulnerable to adversarial attacks, where input data is subtly altered to deceive machine learning models. Malicious actors can make imperceptible changes to images or videos, causing misclassifications or bypassing security measures. These attacks compromise the reliability and robustness of computer vision systems, risking erroneous decisions and security breaches.

DL systems depend on large datasets for training machine learning models, which raises privacy concerns. Collecting, storing, and processing visual data from surveillance cameras or IoT devices risk unauthorized access and data breaches. Another security consideration, like any other software, in computer vision technologies is the vulnerability to exploitation of software flaws or vulnerabilities. Further, integrating computer vision with IoT devices and cloud platforms expands attack vectors and system complexity~\cite{zhou2019discovering}, \cite{surya2016security}, \cite{mishra2021security}. Interconnected systems create dependencies and potential failures, enlarging the attack surface and complicating end-to-end security. 

\subsubsection{Data Security and Privacy}

\begin{enumerate}
    \item \textbf{Data Collection:} Security threats in data collection for DL in the IoV are diverse, encompassing risks such as data tampering, privacy breaches, data poisoning, man-in-the-middle attacks, replay attacks, adversarial attacks, and unauthorized access. Malicious actors may tamper with collected data, compromising its integrity and leading to biased models or incorrect predictions. Unauthorized access to sensitive data threatens privacy, while interception and manipulation during transmission can compromise data integrity. Adversarial attacks deceive DL models by crafting input data. To mitigate these threats, robust encryption, secure communication protocols, access control measures, data authentication techniques, and continuous monitoring are essential. Additionally, secure methods for data collection from vehicles should consider user privacy and data anonymization~\cite{wang2019privacy}.
    
    \item \textbf{Data Storage:} Storage solutions must efficiently manage data while implementing robust security measures to prevent unauthorized access, tampering, or theft. Privacy-preserving techniques are vital to safeguard sensitive information within the data. Balancing data retention policies with regulatory requirements and privacy concerns is crucial. Managing diverse data types and formats from various sources adds complexity, necessitating robust data management strategies and interoperable storage solutions.

    \item  \textbf{Data Transmission:} Issues with data transmission in the IoV pose several challenges, including latency, bandwidth limitations, reliability, security, and protocol compatibility. Latency is critical for real-time decision-making in IoV applications like autonomous driving and traffic management. Bandwidth constraints may limit data transmission, impacting the quality and frequency of updates. Reliable transmission prevents data loss or corruption, especially in dynamic vehicular environments. Security measures must safeguard data from interception or tampering during transmission~\cite{won2015secure},\cite{chatterjee2017puf}. Compatibility issues may arise from integrating heterogeneous data sources or using different communication protocols across IoV components, necessitating interoperability standards. 
    
\end{enumerate}

\subsubsection{Model Security}

\begin{enumerate}
    \item \textbf{Adversarial Attacks:} Adversarial attacks exploit vulnerabilities in DL models by subtly altering input data to mislead the model's decision-making process. In the IoV context, these attacks pose severe risks, such as manipulating traffic signs or causing autonomous vehicles to misinterpret their surroundings. Attackers may perturb sensor data, like GPS signals or camera images, used by DL models in IoV applications, leading to incorrect decisions and potential accidents. Additionally, targeting communication channels between vehicles or infrastructure allows attackers to inject malicious packets or modify data, tricking models into making unsafe predictions or actions. Robust defense mechanisms and rigorous testing are crucial to detect and mitigate such threats, safeguarding the security and reliability of DL models in the IoV.

    \item  \textbf{Model Robustness:} Techniques to increase the robustness of DL models against adversarial and other types of attacks. Model robustness refers to the ability of a DL model to maintain its performance and reliability even when faced with adversarial inputs or perturbations. In the case of IoV, where safety-critical decisions are made based on sensor data processed by DL models, ensuring robustness is crucial to prevent potentially catastrophic outcomes. Adversarial training techniques can train DL models on clean and adversarially perturbed data, allowing the model to learn to recognize and adapt to adversarial inputs, thus enhancing model robustness in IoV. Additionally, incorporating techniques such as input preprocessing, model regularization, and ensembling can help mitigate the impact of adversarial attacks by making the model more resilient to perturbations.
    
    \item  \textbf{Model Verification:} Ensuring the reliability of DL models for the IoV involves rigorous model verification procedures, especially in safety-critical scenarios. Testing encompasses diverse input conditions and edge cases, including exposure to adversarial data, to assess the model's robustness and performance. Validation against ground truth data confirms the accuracy of predictions, while formal verification techniques ensure compliance with safety standards. Regulatory approvals and certifications further validate the model's reliability, enhancing the safety and security of IoV applications like autonomous vehicles and smart transportation systems.

\end{enumerate}

\subsubsection{Deep Learning-Specific Challenges}
\begin{enumerate}
    \item \textit{Explainability:} Developing methods to understand and explain model decisions~\cite{ras2022explainable},\cite{heuillet2021explainability}, \cite{fazi2021beyond} is crucial for diagnosing issues and maintaining trust in automated systems. Without clear insights into how DL models arrive at their decisions or predictions, it becomes challenging to trust their outputs, especially in safety-critical applications like autonomous driving. The opacity of DL models hinders the ability to understand the rationale behind their decisions, making it difficult to diagnose errors, identify biases, or ensure fairness. Moreover, the inability to explain DL model decisions can lead to compliance issues and liability concerns in regulatory or legal contexts~\cite{padovan2023black}. Addressing the challenge of explainability is crucial for enhancing transparency, trustworthiness, and accountability in DL-powered IoV systems, facilitating their adoption and ensuring their safe and ethical deployment in real-world scenarios.
    
    \item \textit{Transfer Learning Security:} Ensuring security when transfer learning techniques~\cite{zhao2017feature},\cite{zhao2019transfer},\cite{cavusoglu2024novel} are used to adapt models trained in one setting to another, which is common in IoV due to the diverse environments and scenarios. While transfer learning offers the advantage of leveraging pre-trained models to enhance learning efficiency and performance, it also introduces security risks. Adversaries could exploit vulnerabilities in transfer learning frameworks to inject malicious data or manipulate model parameters, leading to compromised model integrity and reliability. Moreover, transferring knowledge from source domains to target domains may inadvertently transfer biases or privacy-sensitive information, raising concerns about data leakage and fairness. Securing transfer learning mechanisms in IoV entails implementing robust authentication, encryption, and access control measures to safeguard data integrity and confidentiality during model transfer and adaptation processes. Additionally, continuous monitoring and validation of transferred models are imperative to effectively detect and mitigate potential security threats, ensuring the trustworthiness and resilience of DL-powered IoV systems against adversarial attacks and privacy breaches.
\end{enumerate}

\subsection{System Security in Internet of Vehicles}

One of the primary concerns in IoV is the security of vehicle-to-vehicle (V2V) and vehicle-to-infrastructure (V2I) communication channels. These links enable real-time data exchange for collision avoidance, traffic management, and remote diagnostics but expose vehicles to threats like eavesdropping, spoofing, and tampering~\cite{meneghello2019iot}. Moreover, the growing number of in-vehicle systems and connected services increases potential cyberattack avenues as modern vehicles integrate diverse software ecosystems, including embedded control units, infotainment systems, telematics modules, and wireless interfaces.

IoV systems also face insider risks from compromised or malicious actors with vehicle access. Insider threats involve legitimate users or employees abusing privileges to tamper with settings, steal data, or launch internal network attacks. Privacy concerns arise from the collection and processing of vehicle-related data, highlighting issues of data ownership, consent, and misuse.

\subsubsection{Network Security}
\begin{enumerate}
    \item \textbf{Network Intrusion Detection:} Implementing and maintaining systems to detect and respond to unauthorized access or anomalies in-vehicle networks~\cite{aljarrah2019intrusion}, \cite{kang2016intrusion}.  With the proliferation of interconnected vehicles and infrastructure systems, detecting and mitigating potential threats to network integrity and confidentiality is paramount. Network intrusion detection systems (NIDS) play a crucial role in identifying suspicious activities, such as unauthorized access attempts, malware propagation, and denial-of-service attacks, within IoV environments. Leveraging advanced techniques such as DL and NIDS, it can analyze vast amounts of network traffic data in real time, enabling the detection of anomalies and malicious patterns that traditional methods may overlook. By continuously monitoring network traffic and applying anomaly detection algorithms, NIDS can effectively enhance the security posture of IoV ecosystems, mitigating the risks posed by cyber threats and ensuring the integrity and reliability of vehicle-to-vehicle and vehicle-to-infrastructure communications.
    
    \item \textbf{Secure V2X Communication:} Ensuring the security of vehicle-to-vehicle (V2V) \cite{limbasiya2016secure}, \cite{aliev2020scalable}, \cite{das2022secure} security and vehicle-to-everything (V2X) communications \cite{hasan2020securing}, \cite{qiu2019secure} are fundamental in IoV for cooperative navigation and safety. With the growing connectivity and reliance on vehicles and infrastructure systems, securing V2X communication channels is paramount to thwart potential cyber threats. Secure V2X Communication employs robust encryption, authentication mechanisms, and secure key management protocols to protect sensitive information transmission against eavesdropping, tampering, and unauthorized access. Technologies like blockchain and edge computing enhance resilience and reliability, enabling secure interactions among vehicles, roadside units, and central servers. 
    % Through proactive measures and advanced security solutions, Secure V2X Communication fortifies the overall security posture of IoV environments, fostering trust, privacy, and safety for all stakeholders.

\end{enumerate}

\subsubsection{Identity and Access Management}

\begin{enumerate}
    \item \textbf{Authentication and Authorization:} Vehicle and user identity verification methods are crucial to ensure secure access to specific features or data, forming the foundation of Identity and Access Management (IAM)~\cite{manogaran2023token},~\cite{han2020zero},~\cite{wu2023decentralized}. Authentication validates the identity of entities seeking system access, typically using credentials like usernames, passwords, biometrics, or digital certificates. Following authentication, authorization determines the access level granted to the entity based on predefined permissions and policies. Implementing authentication and authorization mechanisms effectively enforces the principle of least privilege, reducing the risk of unauthorized access and protecting sensitive data from security breaches.
    
    \item \textbf{Digital Certificates:} Digital certificates, provided by trusted Certificate Authorities (CAs), authenticate entities in the IoV~\cite{garcia2018security}. Issued by CAs, these cryptographic credentials bind an individual's or entity's identity to a public key, ensuring the authenticity of digital signatures and enabling secure communication over networks like the Internet. IAM systems use digital certificates to uphold data exchange integrity, confidentiality, and authenticity, bolstering security and facilitating secure resource access. Further, digital certificates enable the deployment of secure protocols like Transport Layer Security (TLS) and Secure Sockets Layer (SSL), safeguarding sensitive information transmitted across communication channels.
\end{enumerate}

To address issues in the IoV ecosystem, stakeholders must adopt a comprehensive software security approach that includes secure development practices, such as coding standards, threat modeling, and code reviews, to identify and fix vulnerabilities early. Cryptographic protections like encryption and secure communication protocols ensure data integrity and confidentiality. Intrusion detection and prevention systems within vehicles and infrastructure enable real-time threat detection and response, while secure over-thin-air (OTA) updates deliver patches efficiently, minimizing exploitation risks. User education initiatives inform about cybersecurity threats and best practices, emphasizing the importance of regular updates. Additionally, enforcing regulatory frameworks and standards ensures industry-wide accountability, transparency, software security, and data protection compliance.

\subsection{Human Aspect}

In the human side of securing tomorrow's smart cities, investigating software security in IoV and DL technologies encompasses several key topics. Firstly, educating users and stakeholders about the potential security risks associated with IoV and DL technologies is crucial~\cite{dekker2021human}. This involves raising awareness about common threats, best practices for secure usage, and the importance of regular software updates and patches. Understanding how human behavior influences the security of IoV and DL systems is essential. This includes studying factors such as user trust, decision-making processes, and the impact of cognitive biases on security-related choices. Examining the privacy implications of IoV and DL technologies is paramount, focusing on user data collection practices, anonymization techniques, and the effectiveness of privacy-preserving measures. Addressing ethical dilemmas from deploying IoV and DL systems involves evaluating algorithmic bias, fairness, accountability, and societal impact. Ensuring compliance with security and privacy regulations is imperative, as well as exploring legal frameworks and their adequacy in addressing emerging challenges. Promoting a human-centric approach to security design integrates user feedback and usability considerations, creating intuitive interfaces and user-friendly features~\cite{erccil2023human}. Lastly, building a skilled workforce capable of addressing evolving cybersecurity threats entails research on effective training methods, skill development programs, and strategies for retaining cybersecurity professionals.

\section{Mitigation Strategies}

Given the critical nature of IoV applications such as autonomous driving and traffic management, ensuring the security and reliability of DL models deployed in these systems is crucial. One effective mitigation strategy involves enhancing the robustness of DL models through rigorous testing and validation procedures~\cite{zhao2023ccprobust},~\cite{liu2021super}. By subjecting the models to diverse and realistic scenarios, researchers can identify and address potential vulnerabilities, such as adversarial attacks and data perturbations, before deploying them in IoV environments. Additionally, incorporating uncertainty estimation techniques~\cite{plasencia2021managing} into DL models can help quantify the confidence levels of model predictions and mitigate the risks associated with uncertainty in real-world scenarios.

Furthermore, adopting privacy-preserving techniques~\cite{kumar2021p2sf} is paramount for safeguarding sensitive data in IoV systems. Differential privacy, homomorphic encryption, and federated learning are among the privacy-preserving approaches that can protect user data privacy while still allowing for collaborative model training and inference in distributed IoV environments. Moreover, implementing robust authentication and access control mechanisms can prevent unauthorized access to IoV systems and mitigate the risks of data breaches and cyber-attacks~\cite{sharma2018blockapp},~\cite{bojjagani2022secure}. IoV stakeholders can effectively mitigate the vulnerabilities introduced by DL techniques and ensure the safety and security of connected vehicles and infrastructure by adopting a multi-layered approach to security, including software and hardware-based defenses.

\section{Future Directions}
Future research in computer vision technologies within the IoT domain reveals several promising directions. Exploring novel techniques for efficient and scalable visual data processing at the edge of IoT networks is crucial. This involves lightweight, energy-efficient algorithms, hardware accelerators, and distributed computing architectures tailored to resource-constrained IoT devices. Advancements in federated learning and edge computing can facilitate collaborative model training and inference across IoT devices, preserving data privacy and reducing bandwidth needs.

Integrating edge computing capabilities with federated learning techniques can enhance the privacy, efficiency, and scalability of IoV systems. By distributing model training and inference tasks across edge devices while preserving data privacy, federated learning enables collaborative intelligence within the IoV ecosystem. Exploration of multi-modal fusion techniques and sensor data integration methods will be crucial for extracting comprehensive insights from heterogeneous IoV data sources. By combining data from various sensors such as cameras, LiDAR, and radar, IoV systems can achieve a more holistic understanding of the surrounding environment and improve decision-making processes.

Further, research into enhancing the security and privacy of computer vision-enabled IoT systems is critical for fostering trust and adoption. Future efforts should focus on developing robust security mechanisms, including secure authentication, encryption, and anomaly detection, to protect against cyber threats and unauthorized access. Additionally, exploring privacy-preserving techniques such as differential privacy, federated learning, and homomorphic encryption can mitigate privacy risks associated with collecting and processing visual data. 

Collaborative efforts between academia, industry, and regulatory bodies are essential for driving the standardization and adoption of IoV technologies. By establishing common protocols, interoperability standards, and best practices, stakeholders can accelerate developing and deploying IoV solutions while ensuring compatibility and reliability across different platforms and vendors.

\section{Conclusion}
In conclusion, our study underscores the critical imperative of prioritizing security in developing Internet of Vehicles (IoV) and Deep Learning (DL) technologies within smart cities. We identify key challenges, including susceptibility to adversarial attacks, data privacy concerns, and the complexity of interconnected systems. Adversarial attacks pose significant risks to DL algorithms, while data collection raises ethical and legal challenges. Integration with IoT devices and cloud platforms adds complexity and additional attack vectors, highlighting the need for comprehensive security measures. By embracing secure development practices, privacy-preserving mechanisms, and collaboration, stakeholders can enhance the resilience and trustworthiness of smart city infrastructure, realizing the full potential of these technologies.

\bibliographystyle{splncs04}
\bibliography{refs}




% LNCS Homepage, \url{http://www.springer.com/lncs}. Last accessed 4
% \end{thebibliography}
\end{document}